\documentstyle[twocolumn,aps,psfig]{revtex}
\begin{document}
\draft
\wideabs{          %Makes a two-column abstract    
\title{Anisotropic effect of field on the orthorhombic-to-tetragonal transition in the %%@
striped cuprate $\rm (La,Nd)_{2-x}Sr_xCuO_4$.}
\author{Z.A. Xu$^{1,*}$, N. P. Ong$^{1}$, T. Noda$^{2}$, H. Eisaki$^{2}$ and S. %%@
Uchida$^{2}$}      
\address{$^1$Joseph Henry Laboratories of Physics, Princeton University, Princeton, New %%@
Jersey 08544}
\address{$^2$School of Frontier Sciences, University of Tokyo,
Yayoi 2-11-16, Bunkyo-ku, Tokyo 113-8656, Japan}
\date{\today}      % Deleting this command produces today's date.

%\newcommand{\ip}[2]{(#1, #2)}
                             % Defines \ip{arg1}{arg2} to mean
                             % (arg1, arg2).

\maketitle                   % Produces the title.

\begin{abstract}
The Nd-doped cuprate $\rm La_{2-y-x}Nd_ySr_xCuO_4$ displays a first-order phase transition %%@
at $T_d$ (= 74 K for $x$=0.10, $y$ = 0.60) to a low-temperature tetragonal (LTT) phase. A %%@
magnetic field $\bf H$ applied $\parallel$ the $a$-axis leads to an {\em increase} in %%@
$T_d$, whereas $T_d$ is decreased when $\bf H\parallel c$.  These effects show that %%@
magnetic ordering involving both Nd and Cu spins plays a key role in driving the LTO-LTT %%@
transition.  Related anisotropic effects are observed in the uniform susceptibility and %%@
the in-plane magnetoresistance. 
\end{abstract}
}				%End of Wideabs command
The observation of static charge and spin modulation in the Nd-doped cuprate $\rm La_{2-x-%%@
y}Nd_ySr_xCuO_4$ \cite{Tranq1,Tranq2} has stimulated considerable interest in stripe %%@
formation in underdoped cuprates. The compound $\rm La_{1.6-x}Nd_{0.4}Sr_xCuO_4$ with $x= %%@
0.12$ undergoes an orthorhombic to tetragonal (LTO-LTT) transition at $T_d$ = 74 K.  %%@
Tranquada {\em et al.} \cite{Tranq1} recently observed weak, static charge and spin %%@
modulations which appear at temperatures below $T_d$.  The modulation periods are close to %%@
$\epsilon$ and 2$\epsilon$ for the charge and spin stripes, respectively, with %%@
$\epsilon\sim x$.  The charge modulation has also been observed by hard $x$-ray scattering %%@
in samples with Sr content $x$ = 0.12 \cite{Zimmermann} and 0.15 \cite{Niemoller}.  %%@
Transport measurements reveal that the $c$-axis resistivity $\rho_c$ increases rapidly at %%@
$T_d$, whereas the Hall coefficient decreases monotonically to a value near zero, implying %%@
quasi-one dimensional transport \cite{Nakamura,Uchida}.  The close similarities between %%@
the static stripe structure and the inelastic peaks observed in Nd-free $\rm La_{2-%%@
x}Sr_xCuO_4$ (LSCO) suggest that weak charge ordering may be a general feature of the %%@
cuprate phase diagram \cite{Yamada}.  Evidence for charged-stripe phase in LSCO (for %%@
$0.07<x<0.12$) was recently derived from the collapse of the NQR (nuclear quadrupole %%@
resonance intensity \cite{Imai}.  Underlying the strong interest in stripes is the %%@
proposal that they are responsible for superconductivity in the cuprates \cite{Kivelson}.

We report an unusual effect of a magnetic field on $T_d$ that highlights the intrinsically %%@
magnetic nature of the LTO-LTT transition. An in-plane field $\bf H\parallel a$ leads to %%@
an increase in $T_d$ while a field $\bf H\parallel c$ decreases $T_d$.  The $T$ dependence %%@
of the uniform susceptibility and in-plane magnetoresistance are also different, depending %%@
on the direction of the field. 

\section{Experimental}
To resolve the MR signal to a few parts per million (ppm), we have calibrated of the %%@
field-sensitivity of the regulating thermometer (Lake Shore Cryogenics cernox sensor).  A %%@
previous calibration (obtained by allowing the He bath to equilibrate under open-loop %%@
control) \cite{Harris} determined that $\Delta\rho/\rho B^2$ = -1.7 and -2.7 ppm, at 100 %%@
and 70 K, respectively.  

\begin{figure}[h]
\centerline{\psfig{figure=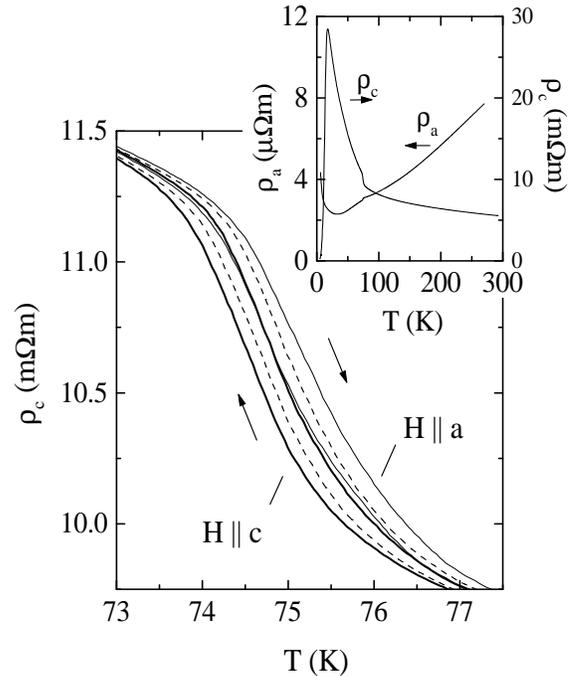,height=3.6in,width=3.0in}}
\caption{(Main Panel) Hysteresis curves of $\rho_c$ vs. $T$ near $T_d$ in $\rm La_{2-y-x} %%@
Nd_ySr_x CuO_4$ (Sample A1, with $x$=0.10).  Curves are measured in zero field (broken %%@
lines), in a fixed 14-T field $\bf H\parallel c$ (heavy solid lines), and in a 14-T field %%@
$\bf H\parallel a$ (light solid lines).  $T_d$ is shifted down (up) if the field is %%@
parallel to the $c$-axis ($a$-axis).  The inset shows the resistivities $\rho_a$ and %%@
$\rho_c$ in A1. }
\label{hyst1}
\end{figure} 
\noindent
We confirmed these numbers by independent tests, and also checked that the sensitivity is %%@
unaffected by the field tilt-angle.  At our largest field (14 T), the error in the set-%%@
point temperature is $<$35 mK, and easily compensated for.  The crystals were cut from %%@
boules grown in an image furnace. Samples A1 and A2 have the same Sr content, $x$ = 0.10, %%@
while Sample B has a higher $x$ (0.15) (the Nd content $y= 0.6$ is the same in all %%@
samples).

\section{Field effect on $T_d$}
In zero field, the $T$ dependences of $\rho_a$ and $\rho_c$ (Fig. \ref{hyst1}, inset) are %%@
similar to those previously observed in crystals of $\rm La_{2-y-x}Nd_ySr_xCuO_4$ with $y$ %%@
= 0.4 and $x$ = 0.12 \cite{Nakamura}.  Whereas $\rho_a$ is only slightly affected by the %%@
LTO-LTT transition, $\rho_c$ increases steeply below $T_d$ until 20 K, where a broad %%@
transition to superconductivity begins. As shown in Fig. \ref{hyst1} (main panel), %%@
$\rho_c$ is hysteretic near $T_d$ between warming and cooling, consistent with a first-%%@
order phase transition.  In an applied field $\bf H$, the hystersis loop is displaced down %%@
in $T$ if $\bf H\parallel c$, but is displaced up if $\bf H\parallel a$ (the upward %%@
displacement is independent of the direction of $\bf H$ within the $ab$ plane).  This %%@
indicates a corresponding shift in the transition temperature $T_d$.  The shift $\Delta %%@
T_d$ is nonlinear in $B^2$. For $\bf H\parallel a$, $\Delta T_d\sim$ +250 mK at 14 T (it %%@
is negative and slightly smaller if $\bf H\parallel c$).  

\begin{figure}[h]
\centerline{\psfig{figure=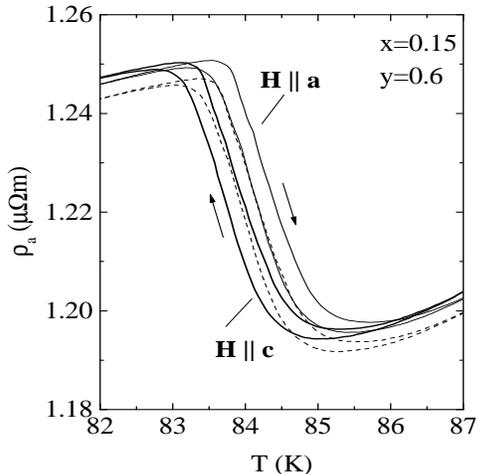,height=2.5in,width=2.5in}}
\caption{Hystereses curves of $\rho_a$ vs. $T$ in Sample B (with $y$=0.6, $x$=0.15) taken %%@
in zero field (broken lines), in a 14-T field $\bf H\parallel c$ (heavy solid lines), and %%@
in a 14-T field $\bf H\parallel a$ (light solid lines).  
}
\label{hyst2}
\end{figure}

In sample B (near optimal doping) the shift $\Delta T_d$ is also similar in magnitude.  In %%@
Fig. \ref{hyst2}, we show the displacements of the hysteresis loop for the in-plane %%@
resistivity $\rho_a$.  The relatively large MR signal raises $\rho_a$ overall by $\sim %%@
0.4\%$ when the field is at 14 T.  Nonetheless, the shifts in $T_d$ (positive when $\bf %%@
H\parallel a$ and negative $\bf c$) are quite apparent. 

\noindent

The unusual shift of $T_d$, together with the steep increase of $\rho_c$ below $T_d$, %%@
leads to a large contribution to the MR observed at fixed temperature (in addition to the %%@
intrinsic MR response).  The profile of $\rho_c$ vs. $T$ suggests that, in the vicinity of %%@
$T_d$, $\partial\rho_c/\partial T$ is most sensitive to $T-T_d$.  Hence, we may write for %%@
the `weak-field' regime (where $\Delta\rho_c\sim B^2$), 
\begin{equation}
\left( \frac{\partial\rho_c}{\partial B^2}\right)_{obs} = \left( %%@
\frac{\partial\rho_c}{\partial B^2}\right)_{int} - \frac{\partial\rho_c}{\partial %%@
T}\frac{\partial T_d}{\partial B^2}.
\label{rhoc}
\end{equation}
The first and second terms on the right are the intrinsic and the $T_d$-shift terms, %%@
respectively.  To demonstrate that both terms exist, we have measured the MR response in %%@
both field geometries (hereafter, we focus on Samples A1 and A2 in which $x$= 0.10).  

As dictated by the sign of $\partial T_d/\partial B^2$, the $T_d$ term adds a spike to the %%@
MR signal that is negative for $\bf H\parallel c$ and positive for $\bf H\parallel a$.
\begin{figure}[h]
\centerline{\psfig{figure=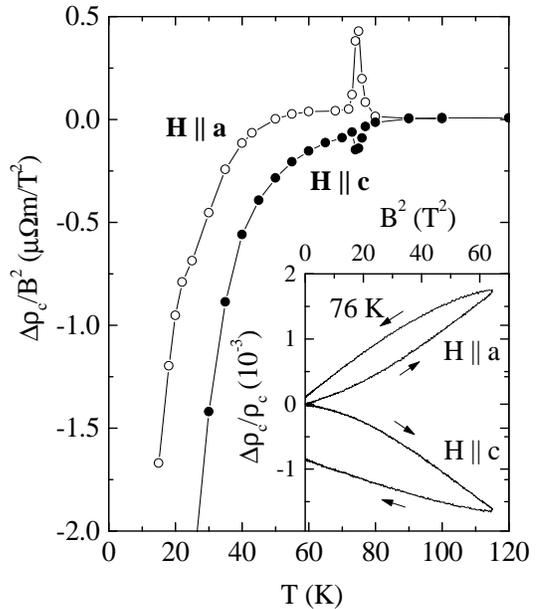,height=3.2in,width=2.8in}}
\caption{(Main Panel) The MR signal $\Delta\rho_c$ (divided by $B^2$) versus $T$ for $\bf %%@
H\parallel c$ (solid circles) and $\bf H\parallel a$ (open) (the latter changes sign near %%@
40 K) in Sample A1.  The MR is expressed as a fractional change in $\rho_c$ at 1 Tesla.  %%@
The spikes at $T_d$ are caused by the $T_d$-shift contribution.  The inset displays %%@
hystereses vs. $H$ with $T$ fixed at 76 K. With the sample prepared on the upper branch of %%@
the $\rho_c-T$ hystesis loop at $H$=0, a large remanent $\Delta\rho_c$ is observed after %%@
field-cycling to 8 T and back to 0, if $\bf H\parallel c$. The remanence is absent if $\bf %%@
H\parallel a$ (see text).
}
\label{mrc}
\end{figure} 
\noindent
This is confirmed by our measurements.  The spikes in the MR curves taken at fixed $T$ %%@
provide an independent check that $T_d$ is shifted up (down) when $\bf H\parallel a$ ($\bf %%@
H\parallel c$).  

Because $\rho_c$ is hysteretic close to $T_d$ (Fig. \ref{hyst1}), it also displays strong %%@
hystereses when $H$ is cycled in the vicinity of $T_d$.  A detailed analysis shows that, %%@
with $T$ fixed, increasing $\bf H \parallel a$ is equivalent to decreasing the reduced %%@
temperature $\epsilon = (T-T_d)/T_d$ in zero field, whereas increasing $\bf H \parallel c$ %%@
increases $\epsilon$.  We start with the system on the upper branch of the hysteretic loop %%@
in Fig. \ref{hyst1} at $H =0$.  If $H$ is increased slowly along the $c$-axis, and then %%@
returned to zero, $\Delta\rho_c$ displays a large negative `remanence' (Fig. \ref{mrc}, %%@
inset).  If instead, $\bf H$ is applied $\bf \parallel a$, the field cycling leads to zero %%@
remanence (the opposite is true on the lower branch).  For the two field geometries, the %%@
distinctive correlation between the remanence (finite or zero) and the starting branch %%@
agrees well with predictions based on the execution of minor hysteresis loops as %%@
$\epsilon$ is cycled in zero field.  It lends further support to the shift in $T_d$ with %%@
field.  The MR signals in the spike region shown in the main panel of Fig. \ref{mrc} %%@
correspond to the initial $\Delta\rho_c$ with the system in the lower (upper) branch for %%@
$\bf H\parallel a$ ($\bf H\parallel c$).

\section{1D Antiferromagnets}
In magnetic systems, a shift in the ordering temperature that changes sign with field %%@
direction is quite rare.  [In 3$D$ antiferromagnets (AF), the applied field invariably %%@
leads to a decrease in the N{\'e}el temperature $T_N$ \cite{Nagamiya}.]  The only %%@
previously known case appears to be quasi-one-dimensional (1D) antiferromagnetic systems, %%@
such as $\rm CsMnCl_3.2H_2O$ and $\rm CsNiCl_3$ \cite{Butterworth,Almond,Jonge}.  These %%@
materials are well-described as 1D Heisenberg antiferromagnets that become 3D-ordered at %%@
low $T$ ($T_N \sim 5 $K) because of a weak interchain coupling.  A field parallel to the %%@
easy axis ($\bf b$) depresses $T_N$ whereas a field perpendicular to $\bf b$ increases %%@
$T_N$.  Calculations \cite{Blume} show that, in a classical 1D antiferromagnet at low $T$, %%@
a magnetic field leads to an increase in the staggered susceptibility $\chi_{\perp}(q)$ %%@
with $q=\pi$.  Villain and Loveluck (VL)\cite{Villain} interpret the enhancement in terms %%@
of the correlation length $\xi$ along the chain.  
\begin{figure}[h]
\centerline{\psfig{figure=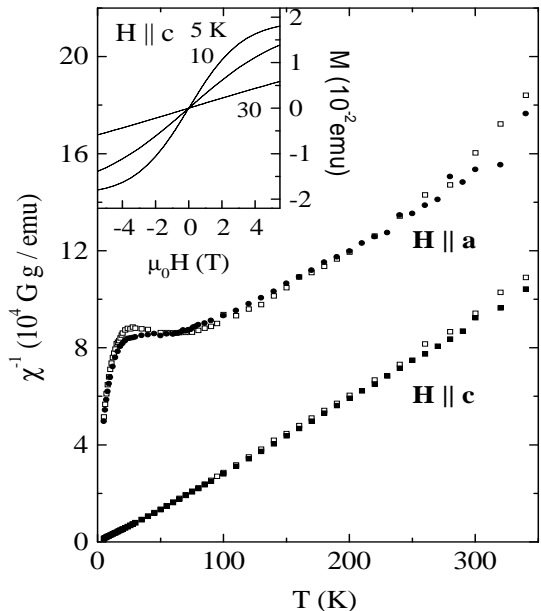,height=3.2in,width=2.8in}}
\caption{The $T$ dependence of the inverse susceptibility $\chi^{-1}$ at 1 Tesla for $\bf %%@
H\parallel a$ and $\bf \parallel c$ in Samples A1 (open) and A2 (closed symbols).  The %%@
inset shows the magnetization $M$ vs. $\bf H\parallel c$ at low $T$ in 2.  The $M-H$ curve %%@
for the $a$-axis geometry is nearly linear even at 5 K.}
\label{chi}
\end{figure} 
\noindent
Suppression of magnetic fluctuations along the field direction has the effect of reducing %%@
by one the number of spin components (from Heisenberg to $XY$, for instance).  This %%@
implies an increase in $\xi$, which in turn enhances $T_N$. Fits using these calculations %%@
account well for the $H$ dependence of $T_N$ observed in several compounds \cite{Jonge}.  %%@
Before applying these ideas to our samples, we discuss the susceptibility measurements.

\section{Susceptibility}
The uniform susceptibility $\chi$ in our samples was measured in a 1-Tesla field using a %%@
SQUID magnetometer.  Figure \ref{chi} compares the $T$ dependence of $\chi_a^{-1}$ and %%@
$\chi_c^{-1}$ obtained with $\bf H\parallel a$ and $\bf H\parallel c$, respectively.  For %%@
$\bf H\parallel c$, $\chi_c^{-1}$ is Curie-like, except that the high-$T$ straight-line %%@
extrapolations show intercepts of 4.9 (8.5) K in Sample A1 (A2).  The slope of $\chi_c^{-%%@
1}$ yields a large effective moment per Nd ion, $p_{eff}= 5.1\; (5.3) \;\mu_B$ (Bohr %%@
magneton).  Our $\chi_c$ is qualitatively similar to the susceptibility in an earlier %%@
report \cite{Tranq2} for $\chi$ measured with $\bf H$ along an unspecified direction.  As %%@
shown in the inset, the magnetization $M$ vs. $\bf H\parallel c$ is linear above 10 K, but %%@
shows incipient ferromagnetic behavior below.  The rapid increase in ordering of the Nd %%@
moments below 3 K has been detected by neutron scattering \cite{Tranq2}.  The large %%@
magnitude of $p_{eff}$ and the apparent ferromagnetic ordering imply that the uniform %%@
susceptibility is dominated by the Nd spins $\bf S$.  Surprisingly, the in-plane %%@
susceptibility $\chi_a$ is qualitatively different in behavior.  Above 100 K, it is %%@
consistent with Curie-Weiss behavior $\chi_a\sim 1/(T+\theta)$ with $\theta\sim$ 180 K.  %%@
Below $T_d$, $\chi_a^{-1}$ displays a levelling off that is reminiscent of the response of %%@
an antiferromagnet.     

The nearly Curie-like behavior of $\chi_c$ implies that the $c$-axis components of the Nd %%@
spins $S_z$ behave as free spins right through $T_d$ (we take $\bf \hat{z}\parallel c$).  %%@
However, the profile of $\chi_a$ shows that the in-plane components $S_x$ and $S_y$ %%@
undergo ordering at $T_d$.  The measurements suggest that $S_x$ and $S_y$ are strongly %%@
coupled to the Cu spins as well as to the phonon mode associated with the tilting of the %%@
octahedra, but $S_z$ is not.  Hence, fluctuations of $S_x$ and $S_y$ strongly influence %%@
magnetic ordering of the Cu spins and the lattice distortion, whereas fluctuations in %%@
$S_z$ do not.  

An attractive picture is that the rapid growth of the staggered susceptibility %%@
$\chi_{xx}({\bf Q})$ is presumably responsible for the LTO-LTT transition [${\bf %%@
Q}=(\pi,\pi)$, nominally]. As in the model of Villain and Loveluck \cite{Villain}, an %%@
external field selectively suppresses fluctuations along the field axis.  When $\bf %%@
H\parallel x$ or $\bf y$, this suppression enhances $\chi_{xx}({\bf Q})$ (and hence %%@
$T_d$).  However, a field $\parallel\bf c$ is ineffectual because fluctuations along $\bf %%@
c$ are irrelevant to the transition at $T_d$. This picture may allow us to relate the %%@
unusual anisotropy of the susceptibility to the striking shift of $T_d$ by an applied %%@
field.  It also emphasizes our key finding that the LTO-LTT transition is intrinsically %%@
magnetic in nature. 

\section{In-plane magnetoresistance}
\begin{figure}[h]
\centerline{\psfig{figure=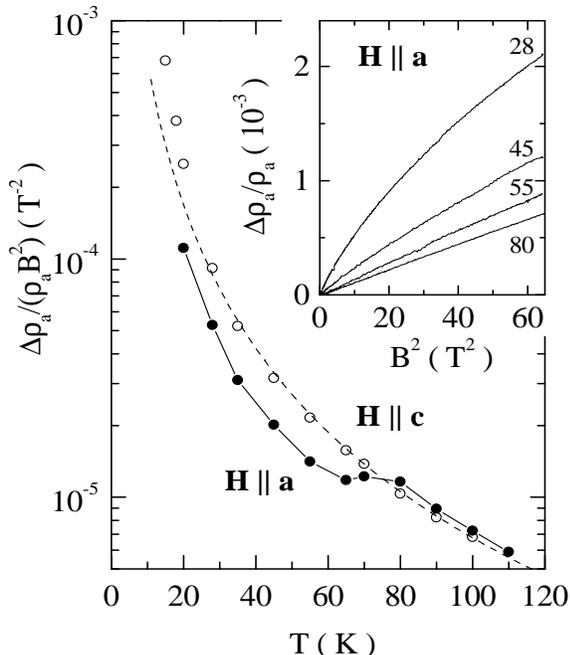,height=3.5in,width=3.0in}}
\caption{The MR signal for the (weak-field) in-plane resistivity $\rho_a$ expressed as the %%@
fractional change in $\rho_a$ divided by $B^2$ (sample A2).  A slight anomaly is observed %%@
in the curve for $\bf H\parallel a$, but not for $\bf H\parallel c$. The broken line is a %%@
$T^2$ fit to the data for $\bf H\parallel c$.  The inset shows that the increase in %%@
$\rho_a$ deviates from $B^2$ below 80 K. 
}
\label{mra}
\end{figure}

The anisotropic nature of the magnetic susceptibility also has striking consequences for %%@
conductance in the plane.  We have measured the in-plane magnetoresistance (current $\bf %%@
J\parallel a$) in Sample A2 for both $\bf H\parallel c$ and $\bf H\parallel a$ (Fig. %%@
\ref{mra}).  Unlike the $c$-axis MR, the in-plane MR is always positive.  Above $T_d$, the %%@
in-plane MR is virtually identical in magnitude for the two field directions, whereas, %%@
below $T_d$, the MR signal is slightly weaker (by $\sim 2$) when $\bf H\parallel a$.  

\noindent

The relatively large MR response (compared with Nd-free LSCO) and its isotropic nature %%@
above $T_d$ imply that it arises entirely from the coupling of $\bf H$ to the spin degrees %%@
of the carriers (as well as to the Nd ions).  Significantly, the MR curve for $\bf %%@
H\parallel c$ passes through $T_d$ without deflection, whereas the curve for $\bf %%@
H\parallel a$ displays a broad anomaly at $T_d$.  The selective sensitivity to the %%@
transition at $T_d$ is consistent with that observed in $\chi$.  The MR data suggest that %%@
the stiffening below $T_d$ of the spin response to a uniform in-plane field also leads to %%@
a decrease in the MR response for the same field direction.  In contrast, the transition %%@
has no effect on $\chi$ or the MR when $\bf H\parallel c$.  Together, the two experiments %%@
reveal rather clearly how spin ordering also influences in-plane charge transport. A %%@
better understanding of how the MR arises may illuminate charge transport in the cuprates.  %%@
We note that the positive sign is opposite to that in conventional spin-mediated MR which %%@
predicts negative MR (the only mechanism that gives a positive MR seems to be from %%@
interaction theory \cite{Lee}).  

In summary, we have described 3 experiments that reveal the magnetic nature of the LTO-LTT %%@
transition in the striped cuprate Nd-doped LSCO.  $T_d$ is shifted up (down) when the %%@
applied field is in-plane (along the $c$-axis).  The uniform susceptibility measured with %%@
$\bf H\parallel a$ shows behavior consistent with AF ordering, but betrays no sign of the %%@
transition when measured with $\bf H\parallel c$.  The same sensitivity to field %%@
orientation is also apparent in the in-plane MR.  The findings all point to a selective %%@
dependence of the LTO-LTT transition to field direction.  We propose that field-%%@
suppression of spin fluctuations along the field axis plays a role similar to that in 1D %%@
Heisenberg systems.  However, the many transport features of the striped phase, such as %%@
the influence of spin ordering on the resistivity, remain to be understood. 
%
%%%   In the acknowledgments, use the following macro  before and instead 
%%%   of  ``Acknoledgments''
%

Discussions with G. Baskaran, D. Huse, B. Keimer, P.A. Lee, and J. Tranquada are %%@
acknowledged. We are especially indebted to B. Keimer for bringing to our attention the 1D %%@
spin-chain results. The work in Princeton is supported by a MRSEC grant (DMR 98-09483) %%@
from the U.S. National Science Foundation.  N.P.O. and S.U. gratefully acknowledge support %%@
from International Joint-Research grants from the New Energy and Industrial Tech. Develop. %%@
Org. (NEDO), Japan.  
\vskip 3mm\noindent
*{\em Permanent address of Z.A.X.} Department of Physics, Zhejiang University, Hangzhou %%@
310027, China.
% now the references. delete or change fake bibitem. 


\begin{thebibliography}{99}
%
\bibitem{Tranq1}  J.M. Tranquada, B.J. Sternlieb, J. D. Axe, Y. Nakamura, and S. Uchida, %%@
Nature {\bf 375}, 561 (1995).
\bibitem{Tranq2}J. M. Tranquada {\em et al.}, Phys. Rev. B {\bf 54}, 7489 (1996).
\bibitem{Zimmermann} M.v. Zimmermann {\em et al.}, Europhys. Lett.  {\bf 41}, 629 (1998).
\bibitem{Niemoller} T. Niem\"{o}ller {\em et al.}, cond-mat/9904383.
\bibitem{Nakamura} Y. Nakamura and S. Uchida, Phys. Rev. B {\bf 46}, 5841 (1992).
\bibitem{Uchida} S. Uchida {\em et al.}, in {\em Physics and Chemistry of Transition Metal %%@
Oxides}, ed. H. Fukuyama and N. Nagaosa, (Springer-Verlag, Heidelberg) 1999, p. 163.
\bibitem{Yamada} K. Yamada et al. Phys. Rev. B {\bf 57}, 6165 (1998).
\bibitem{Imai} 
A.W. Hunt, P. M. Singer, K. R. Thurber, and T. Imai, Phys. Rev. Lett. {\bf 82}, 4300 %%@
(1999).
\bibitem{Singer} 
P. M. Singer, A. W. Hunt, and T. Imai, cond-mat/9906140. 
\bibitem{Kivelson} 
V. J. Emery and S. A. Kivelson, Physica C {\bf 209}, 597 (1993), S. A. Kivelson and V. J. %%@
Emery, {\em Strongly Correlated Electronic Materials: The Los Alamos Symposium 1993} %%@
edited by K.S. Bedell et al. (Addison Wesley, reading, MA), 1994, p. 619. 
\bibitem{Harris} 
J. M. Harris et al., Phys. Rev. Lett. {\bf 75}, 1391 (1995).
\bibitem{Nagamiya} T. Nagamiya, K. Yosida, and R. Kubo, Adv. in Phys. {\bf 4}, 1 (1955).
\bibitem{Jonge} W.J.M. de Jonge, J.P.A. Hijmans, F. Boersma, J.C. Schouten, and K. %%@
Kopinga, Phys. Rev. B {\bf 17}, 2922 (1978).
\bibitem{Butterworth} G. J. Butterworth and J. A. Woollam, Phys. Lett. {\bf 29A}, 259 %%@
(1969).
\bibitem{Almond} D.P. Almond and J. A. Rayne, Phys. Lett. {\bf 55A}, 137 (1975).
\bibitem{Blume} M. Blume, P. Heller, N.A. Lurie, Phys. Rev. B {\bf 11}, 4483 (1975); S. W. %%@
Lovesey and J. M. Loveluck, J. Phys. C {\bf 9}, 3639 (1976).
\bibitem{Villain} J. Villain and J. M. Loveluck, J. de Phys. (Paris) Lett. {\bf 38}, L-77 %%@
(1977).
\bibitem{Lee} 
For a review, see P.A. Lee and T. V. Ramakrishnan, Rev. Mod. Phys. {\bf 57}, 286 (1985). 
\end{thebibliography}
\end{document}